\documentclass{svjour3}                     

\smartqed  
\usepackage{graphicx}
\begin{document}

\title{Block Spin Density Matrix of \\ the Inhomogeneous AKLT Model \thanks{NSF Grant DMS-0503712}}


\titlerunning{Inhomogeneous AKLT Density Matrix}        

\author{Ying Xu \and Hosho Katsura \and Takaaki Hirano \and Vladimir E. Korepin}

\authorrunning{Y. Xu \and H. Katsura \and T. Hirano \and V. E. Korepin} 

\institute{
 Ying Xu \at 
 C.N. Yang Institute for Theoretical Physics \\
 State University of New York at Stony Brook, Stony Brook, NY 11794-3840, USA\\
 \email{yixu@ic.sunysb.edu}
            \and
 Hosho Katsura \at 
 Department of Applied Physics \\ 
 The University of Tokyo, 7-3-1 Hongo, Bunkyo-ku, Tokyo 113-8656, Japan\\ \email{katsura@appi.t.u-tokyo.ac.jp}
            \and
 Takaaki Hirano \at 
 Department of Applied Physics \\ 
 The University of Tokyo, 7-3-1 Hongo, Bunkyo-ku, Tokyo 113-8656, Japan\\ \email{hirano@pothos.t.u-tokyo.ac.jp}           
            \and
           Vladimir E. Korepin \at
 C.N. Yang Institute for Theoretical Physics \\
 State University of New York at Stony Brook, Stony Brook, NY 11794-3840, USA\\
 \email{korepin@insti.physics.sunysb.edu}
}

\date{Received: date / Accepted: date}

\maketitle

\begin{abstract}
We study the inhomogeneous generalization of a 1-dimensional AKLT spin chain model. Spins at each lattice site could be different. Under certain conditions, the ground state of this AKLT model is unique and is described by the Valence-Bond-Solid (VBS) state. We calculate the density matrix of a contiguous block of bulk spins in this ground state. The density matrix is independent of spins outside the block. It is diagonalized and shown to be a projector onto a subspace. We prove that for large block the density matrix behaves as the identity in the subspace. The von Neumann entropy coincides with Renyi entropy and is equal to the saturated value.

\end{abstract}

\section{Introduction}
\label{intro}

Quantum entanglement is a fundamental measure of how much quantum effects we can observe and use to control one quantum system by another, and it is the primary resource in quantum computation and quantum information processing \cite{BD}, \cite{L}. The entanglement of quantum states, particularly related with spin systems has been attracting a great deal of interest, see for example \cite{ABV}, \cite{DHHLB}, \cite{FKR}, \cite{GMC}, \cite{GK}, \cite{GDLL}, \cite{HIZ}, \cite{JK}, \cite{K}, \cite{LRV}, \cite{OL}, 
\cite{PEDC}, \cite{VMC}, \cite{VPC}, 
\cite{W}. Quantum entanglement can be quantified by the von Neumann entropy and Renyi entropy of a subsystem, as discussed in \cite{FKR}, \cite{FIJK}, \cite{FIK}, \cite{IJK}, \cite{KHH}, \cite{VLRK}, \cite{XKHK}. An area law of the von Neumann entropy in harmonic lattice systems has been proposed and studied in \cite{CEP}, \cite{CEPD}, \cite{PEDC}. Entanglement properties also play an important role in condensed matter physics, such as phase transitions \cite{ON}, \cite{OAFF} and macroscopic properties of solids \cite{GRAC}, \cite{V}. 
Extensive research has been undertaken to understand quantum entanglement for correlated electrons, interacting bosons as well as various other systems, see \cite{AFOV}, \cite{AEPW}, \cite{CC}, \cite{CDR}, \cite{CZWZ}, \cite{FK}, \cite{FL}, \cite{FKRHB}, \cite{GLL}, 
\cite{H2}, \cite{HLW}, \cite{ILO}, \cite{KHK}, \cite{KM}, \cite{KP}, \cite{LO}, \cite{LORV}, \cite{LW}, 
\cite{O1}, \cite{O2}, \cite{OLEC}, \cite{PP}, \cite{PHE}, \cite{PS}, \cite{RH}, \cite{V}, \cite{WK}, \cite{WKRL}, \cite{WO}, 
\cite{ZR} 
for reviews and references.  

In this paper we study the inhomogeneous generalization of the AKLT spin chain model. The original homogeneous model was introduced by Affleck, Kennedy, Lieb and Tasaki \cite{AKLT2}, \cite{AKLT1} with all bulk spins being the same. The inhomogeneous generalization was introduced in \cite{KK} in which spins at different lattice sites may take different values. The conditions of the uniqueness of the ground state were discussed and correlation functions in the ground state were obtained. The unique ground state is known as the Valence-Bond-Solid (VBS) state \cite{AAH}, \cite{KK}, which is very important in condensed matter physics. The VBS state is closely related to Laughlin ansatz \cite{LA} and fractional quantum Hall effect \cite{AAH}. It enables us to understand ground state properties of anti-ferromagnetic system with a Haldane gap \cite{H}. Universal quantum computation based on VBS states has also been proposed \cite{VC}.

The density matrix of a contiguous block of bulk spins (we call it \textit{the density matrix} later for short) of the homogeneous AKLT model has been studied extensively in \cite{FKR}, \cite{FM}, \cite{KHH}, \cite{KK}, \cite{VMC}, \cite{XKHK}. It contains information of all correlation functions \cite{AAH}, \cite{JK}, \cite{KHH}, \cite{XKHK}. Moreover, the density matrix was shown \cite{FKR}, \cite{KHH} to be independent of the size of the chain and the location of the block relative to the ends. It was diagonalized in \cite{XKHK} and proved to be a projector onto a subspace. 

It was conjectured in \cite{XKHK} that the structure and properties of the density matrix is generalizable to: (i) the inhomogeneous chain; (ii) higher dimensional lattices; (iii) general graphs. In this paper, we shall prove the first part of the conjecture, i.e. that for the $1$-dimensional inhomogeneous AKLT model. The general version of the inhomogeneous model was first studied in \cite{KK}. In order to write down the Hamiltonian and conditions for the uniqueness of the ground state, we first introduce notations. Consider a system of a linear chain of $N$ bulk spins and two ending spins. By $\vec{S}_{j}$ we denote the vector spin at site $j$ with spin value $S_{j}$. Then we associate a positive integer number to each bond of the lattice and denote by $M_{ij}$ ($M_{ij}=M_{ji}$) the bond number between sites $i$ and $j$. They must be related to bulk spins by the following relation
\begin{eqnarray}
	2S_{j}=M_{j-1,j}+M_{j,j+1} \label{smrela}
\end{eqnarray}
with $2S_{0}=M_{01}$ and $2S_{N+1}=M_{N, N+1}$ for ending spins. The condition for solvability of relation (\ref{smrela}) is
\begin{eqnarray}
	\sum^{N+1}_{j=0}(-1)^{j}S_{j}=0. \label{condition}
\end{eqnarray}
Solution to relation (\ref{smrela}) under condition (\ref{condition}) is 
\begin{eqnarray}
	M_{j,j+1}=2\sum^{j}_{l=0}(-1)^{j-l}S_{l}\geq 1. \label{solution}
\end{eqnarray}
More details can be found in \cite{KK}.

Now we defined the Hamiltonian of the inhomogeneous AKLT model as
\begin{eqnarray}
	H=\sum^{N}_{j=0}\ \sum^{S_{j}+S_{j+1}}_{J=S_{j}+S_{j+1}+1-M_{j,j+1}} C_{J}(j,j+1)P_{J}(j, j+1). \label{hami}
\end{eqnarray}
Here the projector $P_{J}(j, j+1)$ describes interactions between neighboring spins $j$ and $j+1$, which projects the bond spin $\vec{J}_{j, j+1}\equiv\vec{S}_{j}+\vec{S}_{j+1}$ onto the subspace with total spin $J$ ($J=S_{j}+S_{j+1}+1-M_{j,j+1}, \ldots, S_{j}+S_{j+1}$). An explicit expression of $P_{J}(j,j+1)$ was given in \cite{KK}. Coefficient $C_{J}(j,j+1)$ can take arbitrary positive value. This Hamiltonian (\ref{hami}) has a unique ground state (VBS state).

If we consider a block of $L$ contiguous bulk spins as a subsystem, then it is possible to pick up the corresponding terms from (\ref{hami}) which describe interactions within the block. The collection of these terms is referred to as \textit{the block Hamiltonian}:
\begin{eqnarray}
	H_b=\sum^{k+L-2}_{j=k}\ \sum^{S_{j}+S_{j+1}}_{J=S_{j}+S_{j+1}+1-M_{j,j+1}} C_{J}(j,j+1)P_{J}(j, j+1). \label{blockhami}
\end{eqnarray}
Here the block starts from site $k$ and ends at site $k+L-1$. The degeneracy of ground states of the block Hamiltonian (\ref{blockhami}) is $(M_{k-1,k}+1)(M_{k+L-1,k+L}+1)$.

The density matrix $\vec{\rho}$ of the whole chain is a projector onto the unique VBS ground state (see (\ref{pure})). Because of entanglement with spins outside the block, the density matrix $\vec{\rho}_{L}$ of the block as a subsystem (see (\ref{matr})) will no longer be a pure state density matrix as $\vec{\rho}$ is in general. We shall prove that the density matrix $\vec{\rho}_{L}$ is a projector onto a $(M_{k-1,k}+1)(M_{k+L-1,k+L}+1)$-dimensional subspace of the complete Hilbert space associated with the block (see Sections \ref{sec4} and \ref{sec6}). It turns out that the block Hamiltonian $H_{b}$ defines the density matrix $\vec{\rho}_{L}$ completely in the large block limit $L\rightarrow\infty$. The zero-energy ground states of the block Hamiltonian $H_{b}$ span the subspace that the density matrix $\vec{\rho}_{L}$ projects onto. So that $\vec{\rho}_{L}$ can be represented as the zero-temperature limit of the canonical ensemble density matrix defined by $H_{b}$:
\begin{eqnarray}
	\vec{\rho}_{L}=\lim_{\beta\rightarrow +\infty}\frac{e^{-\beta H_{b}}}{Tr\left[e^{-\beta H_{b}}\right]}, \qquad L\rightarrow\infty. \label{enlim}
\end{eqnarray}
In the zero-temperature limit, contributions from excited states of $H_{b}$ all vanish and the right hand side of (\ref{enlim}) turns into a projector onto the ground states of the block Hamiltonian.

As main subjects of the paper, we will construct eigenvectors and derive expressions for corresponding eigenvalues of the density matrix. We will show that the density matrix is a projector.
The paper is divided into three parts:
\begin{enumerate}
	\item We calculate the density matrix using the Schwinger representation (Sections \ref{sec1}, \ref{sec2}).
	\item We construct eigenvectors, derive eigenvalues and prove that the density matrix is a projector. (Sections \ref{sec3}, \ref{sec4}, \ref{sec5})
	\item We investigate the structure of the density matrix in the large block limit. As characteristic functions of quantum entanglement, the von Neumann entropy and the Renyi entropy are obtained in the limit (Section \ref{sec6}).  
\end{enumerate}

\section{Unique Ground State of the Hamiltonian}
\label{sec1}

We start with the ground state of the Hamiltonian (\ref{hami}). It is given in the Schwinger representation by the VBS state \cite{AAH}, \cite{KK}
\begin{eqnarray}
	|\mbox{VBS}\rangle \equiv
 \prod^{N}_{j=0}
\left(a^{\dagger}_{j}b^{\dagger}_{j+1}-b^{\dagger}_{j}a^{\dagger}_{j+1}\right)^{M_{j,j+1}}|\mbox{vac}\rangle, \label{vbs}
\end{eqnarray}
where $a^{\dagger}$, $b^{\dagger}$ are bosonic creation operators and $\left|\mbox{vac}\right\rangle$ is destroyed by any of the annihilation operators $a$, $b$. These operators satisfy $[a_{i}, a^{\dagger}_{j}]=[b_{i}, b^{\dagger}_{j}]=\delta_{ij}$ with all other commutators vanishing. The spin operators are represented as $S^{+}_{j}=a^{\dagger}_{j}b_{j}$, $S^{-}_{j}=b^{\dagger}_{j}a_{j}$, $S^{z}_{j}=\frac{1}{2}(a^{\dagger}_{j}a_{j}-b^{\dagger}_{j}b_{j})$. To reproduce the dimension of the spin-$S_{j}$ Hilbert space at site $j$, an additional constraint on the total boson occupation number is required, namely $\frac{1}{2}(a^{\dagger}_{j}a_{j}+b^{\dagger}_{j}b_{j})=S_{j}$. More details and properties of the VBS state in the Schwinger representation can be found in \cite{AAH}, \cite{A}, \cite{KK}. The pure state density matrix of the VBS ground state (\ref{vbs}) is
\begin{eqnarray}
	\vec{\rho}=\frac{|\mbox{VBS}\rangle\langle \mbox{VBS}|}{\langle \mbox{VBS}|\mbox{VBS}\rangle}. \label{pure}
\end{eqnarray}
Normalization of the VBS state is
\begin{eqnarray}
	\langle \mbox{VBS}|\mbox{VBS}\rangle=\frac{\displaystyle\prod^{N+1}_{j=0}(2S_{j}+1)!}{\displaystyle\prod^{N}_{j=0}(M_{j,j+1}+1)}. \label{normvbs}
\end{eqnarray}
See \cite{XKHK} for more details.

\section{Density Matrix of a Block of Bulk Spins}
\label{sec2}

We take a block of $L$ contiguous bulk spins as a subsystem. In order to calculate the density matrix of the block, it is convenient to introduce a spin coherent state representation. We introduce spinor coordinates
\begin{eqnarray}
	\left(u, v\right)\equiv\left(\cos\frac{\theta}{2}e^{i\frac{\phi}{2}}, \sin\frac{\theta}{2}e^{-i\frac{\phi}{2}}\right), \qquad 0\leq\theta\leq\pi, \quad 0\leq\phi\leq 2\pi. \label{spin}
\end{eqnarray}
Then for a point $\hat{\Omega}\equiv (\sin\theta\cos\phi, \sin\theta\sin\phi, \cos\theta)$ on the unit sphere, the spin-$S$ coherent state is defined as
\begin{eqnarray}
	|\hat{\Omega}\rangle\equiv\frac{\left(ua^{\dagger}+vb^{\dagger}\right)^{2S}}{\sqrt{\left(2S\right)!}}|\mbox{vac}\rangle. \label{cohe}
\end{eqnarray}
Here we have fixed the overall phase (a $U(1)$ gauge degree of freedom) since it has no physical content. Note that (\ref{cohe}) is covariant under $SU(2)$ transforms (see Section \ref{sec6}). The set of coherent states is complete (but not orthogonal) such that \cite{ACGT}, \cite{FM}
\begin{eqnarray}
	\frac{2S+1}{4\pi}\int d\hat{\Omega} |\hat{\Omega}\rangle \langle \hat{\Omega}|=\sum^{S}_{m=-S}|S, m\rangle\langle S, m|=I_{2S+1}, \label{comp}
\end{eqnarray}
where $|S, m\rangle$ denote the eigenstate of $\vec{S}^{2}$ and $S_{z}$, and $I_{2S+1}$ is the identity of the $(2S+1)$-dimensional Hilbert space for spin-$S$. 

Now we calculate the density matrix of a block of $L$ contiguous bulk spins in the VBS state (\ref{vbs}). By definition, this is achieved by taking the pure state density matrix  (\ref{pure}) and tracing out all spin degrees of freedom outside the block:
\begin{eqnarray}
	\vec{\rho}_{L}\equiv Tr_{0, 1, \ldots, k-1, k+L, \ldots, N, N+1}\ \left[\vec{\rho}\right],\qquad 1\leq k,\quad k+L-1\leq N. \label{trac}
\end{eqnarray}
Here the block of length $L$ starts from site $k$ and ends at site $k+L-1$. $\vec{\rho}_{L}$ is no longer a pure state density matrix because of entanglement of the block with the environment (sites outside the block of the spin chain).

Using the coherent state representation (\ref{cohe}) and completeness relation (\ref{comp}), $\vec{\rho}_{L}$ can be written as \cite{KHH}, \cite{XKHK}
\begin{eqnarray}
	&&\vec{\rho}_{L}=
	\label{roug} \\
	&&\frac{
\displaystyle\int\left[\prod^{k-1}_{j=0}\prod^{N+1}_{j=k+L}d\hat{\Omega}_{j}\right]\prod^{k-2}_{j=0}\prod^{N}_{j=k+L}\left[\frac{1}{2}(1-\hat{\Omega}_{j}\cdot\hat{\Omega}_{j+1})\right]^{M_{j,j+1}}
	B^{\dagger}|\mbox{VBS}_{L}\rangle\langle \mbox{VBS}_{L}|B}
	{\displaystyle\left[\prod^{k+L-1}_{j=k}\frac{(2S_{j}+1)!}{4\pi}\right]
\int\left[\prod^{N+1}_{j=0}d\hat{\Omega}_{j}\right]\prod^{N}_{j=0}\left[\frac{1}{2}(1-\hat{\Omega}_{j}\cdot\hat{\Omega}_{j+1})\right]^{M_{j,j+1}}}. \nonumber 
\end{eqnarray}
Here the boundary operator $B$ and block VBS state $\left|\mbox{VBS}_{L}\right\rangle$ are defined as
\begin{eqnarray}
	&&B\equiv \left(u_{k-1}b_{k}-v_{k-1}a_{k}\right)^{M_{k-1,k}}\left(a_{k+L-1}v_{k+L}-b_{k+L-1}u_{k+L}\right)^{M_{k+L-1,k+L}}, \nonumber \\ \label{bope} \\
	&&|\mbox{VBS}_{L}\rangle \equiv
 \prod^{k+L-2}_{j=k}
\left(a^{\dagger}_{j}b^{\dagger}_{j+1}-b^{\dagger}_{j}a^{\dagger}_{j+1}\right)^{M_{j,j+1}}|\mbox{vac}\rangle, \label{vbsl}
\end{eqnarray}
respectively. Note that both $B$ and $|\mbox{VBS}_{L}\rangle$ are $SU(2)$ covariant (see Section \ref{sec6}). After performing integrals over $\hat{\Omega}_{j}$ ($j=0, 1, \ldots, k-2, k+L+1, \ldots, N, N+1$) in the numerator and all integrals in the denominator, the density matrix $\vec{\rho}_{L}$ turns out to be independent of spins outside the block. This property has been proved for homogeneous AKLT models in \cite{FKR}, \cite{KHH}, \cite{XKHK}. Therefore, we can re-label spins within the block for notational convenience. Let $k=1$ and the density matrix takes the form
\begin{eqnarray}
	\vec{\rho}_{L}=\frac{\displaystyle\prod^{L}_{j=0}(M_{j,j+1}+1)}{\displaystyle\prod^{L}_{j=1}(2S_{j}+1)!}\frac{1}{(4\pi)^{2}}
	\int d\hat{\Omega}_{0}d\hat{\Omega}_{L+1}B^{\dagger}|\mbox{VBS}_{L}\rangle\langle \mbox{VBS}_{L}|B \label{matr}
\end{eqnarray}
with
\begin{eqnarray}
&&B^{\dagger}=\left(u^{\ast}_{0}b^{\dagger}_{1}-v^{\ast}_{0}a^{\dagger}_{1}\right)^{M_{0,1}}\left(a^{\dagger}_{L}v^{\ast}_{L+1}-b^{\dagger}_{L}u^{\ast}_{L+1}\right)^{M_{L,L+1}}, \label{bope1} \\
	&&|\mbox{VBS}_{L}\rangle=\prod^{L-1}_{j=1}
\left(a^{\dagger}_{j}b^{\dagger}_{j+1}-b^{\dagger}_{j}a^{\dagger}_{j+1}\right)^{M_{j,j+1}}|\mbox{vac}\rangle. \label{vbsl1}	
\end{eqnarray}
The remaining two integrals of (\ref{matr}) can be performed, but we keep its present form for later use.

\section{Ground States of the Block Hamiltonian}
\label{sec3}

The block Hamiltonian (\ref{blockhami}) with the re-labeling $k=1$ reads
\begin{eqnarray}
	H_b=\sum^{L-1}_{j=1}\ \sum^{S_{j}+S_{j+1}}_{J=S_{j}+S_{j+1}+1-M_{j,j+1}} C_{J}(j,j+1)P_{J}(j, j+1). \label{blockhamire}
\end{eqnarray}
Now we define a set of operators covariant under $SU(2)$:
\begin{eqnarray}
 A^{\dagger}_{J}\equiv
\left(ua^{\dagger}_{1}+vb^{\dagger}_{1}\right)^{J_{-}+J}  \left(a^{\dagger}_{1}b^{\dagger}_{L}-b^{\dagger}_{1}a^{\dagger}_{L}\right)^{J_{+}-J}
\left(ua^{\dagger}_{L}+vb^{\dagger}_{L}\right)^{-J_{-}+J}, \label{aope}
\end{eqnarray}
where $J_{-}\equiv\frac{1}{2}(M_{01}-M_{L,L+1})$, $J_{+}\equiv\frac{1}{2}(M_{01}+M_{L,L+1})$ and $|J_{-}|\leq J\leq J_{+}$. These operators (\ref{aope}) act on the direct product of Hilbert spaces associated with site $1$ and site $L$. Then the set of ground states of (\ref{blockhamire}) can be chosen as
\begin{eqnarray}
	|\mbox{G}; J, \hat{\Omega}\rangle \equiv
A^{\dagger}_{J}|\mbox{VBS}_{L}\rangle, \qquad J=|J_{-}|, \ldots, J_{+}. \label{eige}
\end{eqnarray}
Any state $|\mbox{G}; J, \hat{\Omega}\rangle$ of this set depending on a discrete parameter $J$ and a continuous unit vector $\hat{\Omega}$ is a zero-energy ground state of (\ref{blockhamire}). To prove this we need only to verify for any site $j$ and bond $(j, j+1)$: (i) the total power of $a^{\dagger}_{j}$ and $b^{\dagger}_{j}$ is $2S_{j}$, so that we have spin-$S_{j}$ at site $j$; (ii) $-\frac{1}{2}(M_{j-1,j}+M_{j+1,j+2})\leq J^{z}_{j,j+1}\equiv S^{z}_{j}+S^{z}_{j+1}\leq \frac{1}{2}(M_{j-1,j}+M_{j+1,j+2})$ by a binomial expansion, so that the maximum value of the bond spin $J_{j,j+1}$ is $\frac{1}{2}(M_{j-1,j}+M_{j+1,j+2})=S_{j}+S_{j+1}-M_{j,j+1}$ (from $SU(2)$ invariance, see \cite{AAH}). Therefore, the state $|\mbox{G}; J, \hat{\Omega}\rangle$ defined in (\ref{eige}) has spin-$S_{j}$ at site $j$ and no projection onto the $J_{j, j+1}>S_{j}+S_{j+1}-M_{j,j+1}$ subspace for any bond.

The set of states $\{|\mbox{G}; J, \hat{\Omega}\rangle\}$ with the same $J$ value are not orthogonal. We also introduce alternatively an orthogonal basis in description of the degenerate zero-energy ground states of (\ref{blockhamire}). For notational convenience, we define 
\begin{eqnarray}
	X_{JM}\equiv\frac{u^{J+M}v^{J-M}}{\sqrt{(J+M)!(J-M)!}}, \qquad
	\psi^{\dagger}_{Sm}\equiv\frac{(a^{\dagger})^{S+m}(b^{\dagger})^{S-m}}{\sqrt{(S+m)!(S-m)!}}. \label{2def}
\end{eqnarray}
These two variables transform conjugately with respect to one another under $SU(2)$ (see Section \ref{sec6}). $X_{JM}$ has the following orthogonality relation
\begin{eqnarray}
	\int d\hat{\Omega}X^{\ast}_{JM}X_{JM^{\prime}}=\frac{4\pi}{(2J+1)!}\delta_{MM^{\prime}}. \label{orth}
\end{eqnarray}
$\psi^{\dagger}_{Sm}$ is a spin state creation operator such that
\begin{eqnarray}
	\psi^{\dagger}_{Sm}|\mbox{vac}\rangle=|S, m\rangle. \label{crea1}
\end{eqnarray}
The operator $A^{\dagger}_{J}$ defined in (\ref{aope}) can be expanded as (see \cite{Ha}, \cite{XKHK})
\begin{eqnarray}
	A^{\dagger}_{J}&=&\sqrt{\frac{(J_{+}+J+1)!(J_{-}+J)!(J_{+}-J)!(-J_{-}+J)!}{2J+1}} 
	\sum^{J}_{M=-J}X_{JM} \label{aexp} \\
	&&\cdot\sum^{m_{1}+m_{L}=M}_{m_{1}, m_{L}}(\frac{1}{2}M_{01}, m_{1}; \frac{1}{2}M_{L,L+1}, m_{L}|J, M)\ \psi^{\dagger}_{\frac{1}{2}M_{01},m_{1}}\otimes\psi^{\dagger}_{\frac{1}{2}M_{L,L+1},m_{L}}, \nonumber
	\label{expandAJ}
\end{eqnarray}
where $(\frac{1}{2}M_{01}, m_{1}; \frac{1}{2}M_{L,L+1}, m_{L}|J, M)$ are the Clebsch-Gordan coefficients. Note that $\psi^{\dagger}_{\frac{1}{2}M_{01},m_{1}}$ and $\psi^{\dagger}_{\frac{1}{2}M_{L,L+1},m_{L}}$ are defined in the Hilbert spaces of spins at site $1$ and site $L$, respectively. We realize that the particular form of the sum over $m_{1}$ and $m_{L}$ in (\ref{aexp}) can be identified as a single spin state creation operator
\begin{eqnarray}
	\Psi^{\dagger}_{JM}\equiv\sum^{m_{1}+m_{L}=M}_{m_{1}, m_{L}}(\frac{1}{2}M_{01}, m_{1}; \frac{1}{2}M_{L,L+1}, m_{L}|J, M)\ \psi^{\dagger}_{\frac{1}{2}M_{01},m_{1}}\otimes\psi^{\dagger}_{\frac{1}{2}M_{L,L+1},m_{L}}.\nonumber \\ \label{crea2}
\end{eqnarray}
This operator $\Psi^{\dagger}_{JM}$ acts on the direct product of two Hilbert spaces of spins at site $1$ and site $L$. It has the property that
\begin{eqnarray}
	\Psi^{\dagger}_{JM}|\mbox{vac}\rangle_{1}\otimes|\mbox{vac}\rangle_{L}=|J, M\rangle_{1,L}. \label{crea3}
\end{eqnarray}
If we define a set of \textit{degenerate VBS states} $\{|\mbox{VBS}_L(J,M)\rangle\}$ such that
\begin{eqnarray}
	|\mbox{VBS}_L(J,M)\rangle\equiv \Psi^{\dagger}_{JM}|\mbox{VBS}_L\rangle, \quad J=|J_{-}|,...,J_{+}, \quad M=-J, ...,J, \label{devb}
\end{eqnarray}
then these $(M_{01}+1)(M_{L,L+1}+1)$ states (\ref{devb}) are not only linearly independent but also mutually orthogonal (see Appendix \ref{secA}). Furthermore, any ground state $|\mbox{G}; J, \hat{\Omega}\rangle$ can be written as a linear superposition over these degenerate VBS states as
\begin{eqnarray}
	|\mbox{G};J,\hat{\Omega}\rangle&=&\sqrt{\frac{(J_{+}+J+1)!(J_{-}+J)!(J_{+}-J)!(-J_{-}+J)}{2J+1}}
	\nonumber \\
	&&\cdot\sum^{J}_{M=-J}X_{JM}|\mbox{VBS}_{L}(J, M)\rangle,
	\label{line}
\end{eqnarray}
and \textit{vice versa}.
More details can be found in \cite{Ha}, \cite{XKHK}.

Therefore, as seen from (\ref{line}), the rank of set of states $\{|\mbox{G}; J, \hat{\Omega}\rangle\}$ with the same $J$ value is $2J+1$ and the total number of linearly independent states of the set $\{|\mbox{G}; J, \hat{\Omega}\rangle\}$ is $\sum^{J_{+}}_{J=|J_{-}|}(2J+1)=(M_{01}+1)(M_{L,L+1}+1)$, which is exactly the degeneracy of the ground states of (\ref{blockhamire}). So that $\{|\mbox{G}; J, \hat{\Omega}\rangle\}$ forms a complete set of zero-energy ground states. The orthogonal set $\{|\mbox{VBS}_L(J,M)\rangle\}$ differs from $\{|\mbox{G}; J, \hat{\Omega}\rangle\}$ by a change of basis, so that it also forms a complete set of zero-energy ground states.

\section{Diagonalization of the Density Matrix}
\label{sec4}

It was shown in \cite{XKHK} for the homogeneous AKLT model that the density matrix is a projector onto a subspace. This statement is also true for the inhomogeneous model.

Now we propose a theorem on the eigenvectors of the density matrix (\ref{matr}). The explicit construction of eigenvectors yields a direct diagonalization of the density matrix. The set of eigenvectors also spans the subspace that the density matrix projects onto.

\begin{theorem}
	Eigenvectors of the density matrix $\vec{\rho}_{L}$ (\ref{matr}) with non-zero eigenvalues are given by the set $\{|\mbox{G}; J, \hat{\Omega}\rangle\}$ (\ref{eige}), or, equivalently, by the set $\{|\mbox{VBS}_L(J,M)\rangle\}$ (\ref{devb}). \\ i.e. they are zero-energy ground states of the block Hamiltonian $H_{b}$ (\ref{blockhamire}).
\end{theorem}

We prove the theorem by re-writing the density matrix $\vec{\rho}_{L}$ (\ref{matr}) as a projector in diagonal form onto the orthogonal degenerate VBS states $\{|\mbox{VBS}_{L}(J, M)\rangle\}$ introduced in (\ref{devb}).

Take expression (\ref{matr}) and integrate over $\hat{\Omega}_{0}$ and $\hat{\Omega}_{L+1}$ using binomial expansions and $\int^{1}_{-1}dx (1+x)^{m}(1-x)^{n}=\frac{m!n!}{(m+n+1)!}2^{m+n+1}$. Then we have
\begin{eqnarray}
	\vec{\rho}_{L}&=&	\frac{\displaystyle\prod^{L-1}_{j=1}(M_{j,j+1}+1)}{\displaystyle\prod^{L}_{j=1}(2S_{j}+1)!}\sum^{M_{01}}_{p=0}\sum^{M_{L,L+1}}_{q=0}\left(\begin{array}{c}M_{01}\\p\end{array}\right)\left(\begin{array}{c}M_{L,L+1}\\q\end{array}\right)
\nonumber \\
&&(b^{\dagger}_{1})^{p}(a^{\dagger}_{1})^{M_{01}-p}(a^{\dagger}_{L})^{q}(b^{\dagger}_{L})^{M_{L,L+1}-q}|\mbox{VBS}_{L}\rangle
\nonumber \\
&&\langle\mbox{VBS}_{L}|(b_{L})^{M_{L,L+1}-q}(a_{L})^{q}(a_{1})^{M_{01}-p}(b_{1})^{p}. \label{treat1}
\end{eqnarray}
The particular combinations of bosonic operators appeared in (\ref{treat1}) are recognized up to a constant as spin creation operators $\psi^{\dagger}_{\frac{1}{2}M_{01},\frac{1}{2}M_{01}-p}$ and $\psi^{\dagger}_{\frac{1}{2}M_{L,L+1},q-\frac{1}{2}M_{L,L+1}}$ (see (\ref{2def})) at site $1$ and site $L$, respectively. They commute with all bond operators $\left(a^{\dagger}_{j}b^{\dagger}_{j+1}-b^{\dagger}_{j}a^{\dagger}_{j+1}\right)^{M_{j,j+1}}$, so that we could simplify the right hand side of (\ref{treat1}) using definition (\ref{crea2}) and the following identity:
\begin{eqnarray}
&&\sum^{M_{01}}_{p=0}\sum^{M_{L,L+1}}_{q=0}\psi^{\dagger}_{\frac{1}{2}M_{01},\frac{1}{2}M_{01}-p}\otimes\psi^{\dagger}_{\frac{1}{2}M_{L,L+1},q-\frac{1}{2}M_{L,L+1}}|\mbox{vac}\rangle_{1,L}
\nonumber \\
&&{}_{1,L}\langle\mbox{vac}|\psi_{\frac{1}{2}M_{01},\frac{1}{2}M_{01}-p}\otimes\psi_{\frac{1}{2}M_{L,L+1},q-\frac{1}{2}M_{L,L+1}} \nonumber \\
&=&\sum^{M_{01}}_{p=0}\sum^{M_{L,L+1}}_{q=0}|\frac{1}{2}M_{01},\frac{1}{2}M_{01}-p\rangle_{1}\langle\frac{1}{2}M_{01},\frac{1}{2}M_{01}-p| \nonumber \\
&&\otimes|\frac{1}{2}M_{L,L+1},q-\frac{1}{2}M_{L,L+1}\rangle_{L}\langle\frac{1}{2}M_{L,L+1},q-\frac{1}{2}M_{L,L+1}| \nonumber \\
&=&\sum^{J_{+}}_{J=|J_{-}|}\sum^{J}_{M=-J}|J,M\rangle_{1,L}\langle J,M| \nonumber \\
&=&\sum^{J_{+}}_{J=|J_{-}|}\sum^{J}_{M=-J}\Psi^{\dagger}_{JM}|\mbox{vac}\rangle_{1,L}\langle \mbox{vac}|\Psi_{JM}.
\end{eqnarray}
The resultant final form of density matrix $\vec{\rho}_{L}$ is
\begin{eqnarray}
	&&\vec{\rho}_{L}=\frac{\displaystyle\prod^{L-1}_{j=1}(M_{j,j+1}+1)}{\displaystyle\prod^{L}_{j=1}(2S_{j}+1)!}M_{01}!M_{L,L+1}!\sum^{J_{+}}_{J=|J_{-}|}\sum^{J}_{M=-J}\Psi^{\dagger}_{JM}|\mbox{VBS}_{L}\rangle\langle \mbox{VBS}_{L}|\Psi_{JM} \label{fipr} \nonumber \\
&&=\frac{\displaystyle\prod^{L-1}_{j=1}(M_{j,j+1}+1)}{\displaystyle\prod^{L}_{j=1}(2S_{j}+1)!}M_{01}!M_{L,L+1}!\sum^{J_{+}}_{J=|J_{-}|}\sum^{J}_{M=-J}|\mbox{VBS}_{L}(J, M)\rangle\langle \mbox{VBS}_{L}(J, M)|. \nonumber \\ \label{finalrho}
\end{eqnarray}

The set of degenerate VBS states $\{|\mbox{VBS}_{L}(J, M)\rangle\}$ with $J=|J_{-}|,\ldots,J_{+}$ and $M=-J,\ldots,J$ forms an orthogonal basis (see Appendix \ref{secA}). These $(M_{01}+1)(M_{L,L+1}+1)$ states also forms a complete set of zero-energy ground states of the block Hamiltonian (\ref{blockhamire}). So that in expression (\ref{finalrho}) we have re-written the density matrix as a projector in diagonal form over an orthogonal basis. Each degenerate VBS state $|\mbox{VBS}_{L}(J, M)\rangle$ is an eigenvector of the density matrix, so as any of the state $|\mbox{G}; J, \hat{\Omega}\rangle$ (because of the degeneracy of corresponding eigenvalues of the density matrix, see Section \ref{sec5}). Thus we have proved \textbf{Theorem 1}.

\section{Eigenvalues of the Density Matrix}
\label{sec5}

Given the diagonalized form (\ref{finalrho}), eigenvalues of the density matrix $\vec{\rho}_{L}$ can be derived from normalization of degenerate VBS states. We obtain an explicit expression for eigenvalues in terms of Wigner $3j$-symbols in this section. 

Using expansion (\ref{line}) and orthogonality (\ref{orth}), normalization of $|\mbox{VBS}_{L}(J,M)\rangle$ can be calculated from integrating the inner product of $|\mbox{G};J,{\hat\Omega}\rangle$ with itself over the unit vector $\hat{\Omega}$ such that
\begin{eqnarray}
 &&\frac{1}{4\pi}\int d{\hat \Omega}\langle {\rm G};J,{\hat \Omega}|{\rm G};J,{\hat \Omega}\rangle  \label{novb} \\
&=& \frac{(J_{+}+J+1)!(J_{-}+J)!(J_{+}-J)!(-J_{-}+J)!}{(2J+1)!}\langle \mbox{VBS}_L(J,M)|\mbox{VBS}_L(J,M)\rangle.\nonumber
\end{eqnarray}
This expression (\ref{novb}) also states that normalization of the degenerate VBS state is independent of $\hat{\Omega}$ and/or $M$ (see Appendix \ref{secA} for another proof).

Let's consider the integral involved in (\ref{novb}). Using coherent state representation (\ref{cohe}) and completeness relation (\ref{comp}) as before, we obtain
\begin{eqnarray}
&& \frac{1}{4\pi}\int d{\hat\Omega}\langle\mbox{G};J,{\hat\Omega}|\mbox{G};J,{\hat\Omega}\rangle
\label{inne} \\
&=& \frac{1}{4\pi}\left[\prod^{L}_{j=1}\frac{(2S_{j}+1)!}{4\pi}\right] \int d{\hat\Omega} \int \left[\prod^{L}_{j=1} d{\hat\Omega}_{j}\right] \prod^{L-1}_{j=1}\left[\frac{1}{2}(1-\hat{\Omega}_{j}\cdot\hat{\Omega}_{j+1})\right]^{M_{j,j+1}} \nonumber \\
&\cdot&\left[\frac{1}{2}(1-\hat{\Omega}_{1}\cdot\hat{\Omega}_{L})\right]^{J_{+}-J}
\left[\frac{1}{2}(1+\hat{\Omega}_{1} \cdot\hat{\Omega})\right]^{J_{-}+J}
\left[\frac{1}{2}(1+\hat{\Omega} \cdot\hat{\Omega}_{L})\right]^{-J_{-}+J}. \nonumber
\end{eqnarray}
Now we expand $\left[\frac{1}{2}(1-\hat{\Omega}_i \cdot\hat{\Omega}_j)\right]^{M_{ij}}$ in terms of spherical harmonics as in \cite{FM}, \cite{KHH}, \cite{XKHK}
\begin{eqnarray}
\left[\frac{1}{2}(1-\hat{\Omega}_i \cdot\hat{\Omega}_j)\right]^{M_{ij}}
=\frac{4\pi}{M_{ij}+1}\sum^{M_{ij}}_{l=0}\lambda(l,M_{ij})\sum^{l}_{m=-l} Y_{lm}(\hat{\Omega}_{i})
Y^{\ast}_{lm}(\hat{\Omega}_{j}) \label{expansion}
\end{eqnarray}
with
\begin{equation}
\lambda(l,M_{ij})=\frac{(-1)^l M_{ij}!(M_{ij}+1)!}{(M_{ij}-l)!(M_{ij}+l+1)!}. \label{lambdalm}
\end{equation}
Then integrate over ${\hat\Omega}$ and from ${\hat\Omega}_2$ to ${\hat\Omega}_{L-1}$, the right hand side of (\ref{inne}) is equal to
\begin{eqnarray}
&& \frac{4\pi \displaystyle\prod^{L}_{j=1}(2S_{j}+1)!}{\left[\displaystyle\prod^{L-1}_{j=1}(M_{j,j+1}+1)\right](J_{-}+J+1)(J_{+}-J+1)(-J_{-}+J+1)} \nonumber \\
&\cdot&\sum^{M_{<}}_{l} \sum^{J_{+}-J}_{l_{\alpha}=0} \sum^{J_{<}}_{l_{\beta}=0}
\sum^{l}_{m=-l} \sum^{l_{\alpha}}_{m_{\alpha}=-l_{\alpha}} \sum^{l_{\beta}}_{m_{\beta}=-l_{\beta}}
\left[\prod^{L-1}_{j=1}\lambda(l,M_{j,j+1})\right] \nonumber \\
&\cdot&\lambda(l_{\alpha},J_{+}-J) \lambda(l_{\beta},J_{-}+J) \lambda(l_{\beta},-J_{-}+J)
\int d{\hat\Omega}_{1} \int d{\hat\Omega}_{L}
\nonumber \\
&\cdot& Y_{l,m}({\hat\Omega}_{1})Y_{l_{\alpha},m_{\alpha}}({\hat\Omega}_{1})Y_{l_{\beta},m_{\beta}}({\hat\Omega}_1)
Y^{\ast}_{l,m}({\hat\Omega}_{L})Y^{\ast}_{l_{\alpha},m_{\alpha}}({\hat\Omega}_{L})Y^{\ast}_{l_{\beta},m_{\beta}}({\hat\Omega}_{L}).\label{pre1}
\end{eqnarray}
Where we have $M_{<}\equiv min\{M_{j,j+1},j=1,\ldots,L-1\}$ and $J_{<}\equiv min\{J_{-}+J, -J_{-}+J\}$, both being the minimum of the corresponding set. Now we carry out remaining integrals in (\ref{pre1}) using
\begin{eqnarray}
&&\int d{\hat\Omega} Y_{l,m}({\hat\Omega})Y_{l_{\alpha},m_{\alpha}}({\hat\Omega})Y_{l_{\beta},m_{\beta}}({\hat\Omega})
\nonumber \\
&=&\sqrt{\frac{(2l+1)(2l_{\alpha}+1)(2l_{\beta}+1)}{4\pi}}
\left(\begin{array}{c c c}
l & l_{\alpha} & l_{\beta} \\
0   &  0  & 0 
\end{array}\right)
\left(\begin{array}{c c c}
l & l_{\alpha} & l_{\beta} \\
m & m_{\alpha} & m_{\beta} 
\end{array}\right). \label{wign}
\end{eqnarray}
The result after integration can be further simplified by applying the following orthogonality relation
\begin{eqnarray}
	\sum_{m,m_{\alpha}}(2l_{\beta}+1)
	\left(\begin{array}{c c c}
	l & l_{\alpha} & l_{\beta} \\
	m & m_{\alpha} & m_{\beta} 
	\end{array}\right)
	\left(\begin{array}{c c c}
	l & l_{\alpha} & l^{\prime}_{\beta} \\
	m & m_{\alpha} & m^{\prime}_{\beta} 
	\end{array}\right)=\delta_{l_{\beta}l^{\prime}_{\beta}} \delta_{m_{\beta}m^{\prime}_{\beta}}, \label{3jor}
\end{eqnarray}
where $\left(\begin{array}{c c c}
l & l_{\alpha} & l_{\beta} \\
m & m_{\alpha} & m_{\beta} 
\end{array}\right)$, \textit{etc.} are the Wigner $3j$-symbols.

So that finally expression (\ref{pre1}) is equal to
\begin{eqnarray}
&&\frac{\displaystyle\prod^{L}_{j=1}(2S_{j}+1)!}{\left[\displaystyle\prod^{L-1}_{j=1}(M_{j,j+1}+1)\right](J_{-}+J+1)(J_{+}-J+1)(-J_{-}+J+1)} \nonumber \\
&\cdot&\sum^{M_{<}}_{l} \sum^{J_{+}-J}_{l_{\alpha}=0} \sum^{J_{<}}_{l_{\beta}=0}
\left[\prod^{L-1}_{j=1}\lambda(l,M_{j,j+1})\right]
\lambda(l_{\alpha},J_{+}-J) \lambda(l_{\beta},J_{-}+J) \lambda(l_{\beta},-J_{-}+J)
\nonumber \\
&\cdot&(2l+1)(2l_{\alpha}+1)(2l_{\beta}+1)
\left(\begin{array}{c c c}
l & l_{\alpha} & l_{\beta} \\
0   &  0  & 0 
\end{array}\right)^{2}. \label{renovb}
\end{eqnarray}
The explicit value of $\left(\begin{array}{c c c}
l & l_{\alpha} & l_{\beta} \\
0   &  0  & 0 
\end{array} \right)$ is given by
\begin{eqnarray}
&&\left(\begin{array}{c c c}
l & l_{\alpha} & l_{\beta} \\
0   &  0  & 0 
\end{array} \right) \label{3jva} \\
&=&(-1)^{g} \sqrt{\frac{(2g-2l)!(2g-2l_{\alpha})!(2g-2l_{\beta})!}{(2g+1)!}}\frac{g!}{(g-l)!(g-l_{\alpha})!(g-l_{\beta})!},
\nonumber
\end{eqnarray}
if $l+l_{\alpha}+l_{\beta}=2g$ ($g \in \mbox{\textbf{N}}$), otherwise zero. 

Combining results of (\ref{fipr}), (\ref{novb}) and (\ref{renovb}), we arrive at the following theorem on eigenvalues:
\begin{theorem}
Eigenvalues $\Lambda(J)$ $(J=|J_{-}|,\ldots,J_{+})$ of the density matrix are independent of $\hat{\Omega}$ and/or $M$ in defining eigenvectors (see (\ref{eige}) and (\ref{devb})). An explicit expression is given by the following triple sum
	\begin{eqnarray}
&&\Lambda(J)		=\frac{\displaystyle\prod^{L-1}_{j=1}(M_{j,j+1}+1)}{\displaystyle\prod^{L}_{j=1}(2S_{j}+1)!}{M_{01}!M_{L,L+1}!}\langle\mbox{VBS}_{L}(J,M)|\mbox{VBS}_{L}(J,M)\rangle \label{eivaex} \\
&&=\frac{(2J+1)!M_{01}!M_{L,L+1}!}{(J_{+}+J+1)!(J_{-}+J+1)!(J_{+}-J+1)!(-J_{-}+J+1)!} \nonumber \\
&\cdot&\sum^{M_{<}}_{l} \sum^{J_{+}-J}_{l_{\alpha}=0} \sum^{J_{<}}_{l_{\beta}=0}
\left[\prod^{L-1}_{j=1}\lambda(l,M_{j,j+1})\right]
\lambda(l_{\alpha},J_{+}-J) \lambda(l_{\beta},J_{-}+J) \lambda(l_{\beta},-J_{-}+J)
\nonumber \\
&\cdot&(2l+1)(2l_{\alpha}+1)(2l_{\beta}+1)
\left(\begin{array}{c c c}
l & l_{\alpha} & l_{\beta} \\
0   &  0  & 0 
\end{array}\right)^{2}. \nonumber
	\end{eqnarray}
\end{theorem}

We shall emphasize at this point that given eigenvalues (\ref{eivaex}), both von Neumann entropy 
\begin{eqnarray}
	S_{v.N}=-Tr\left[\vec{\rho_{L}}\ln{\vec{\rho_{L}}}\right]=-\sum^{J_{+}}_{J=|J_{-}|}(2J+1)\Lambda(J)\ln\Lambda(J) \label{vonn}
\end{eqnarray}
and Renyi entropy 
\begin{eqnarray}
	S_{R}=\frac{1}{1-\alpha}\ln \left\{Tr\left[\vec{\rho}^{\alpha}_{L}\right]\right\}=\frac{1}{1-\alpha}\ln\left\{\sum^{J_{+}}_{J=|J_{-}|}(2J+1)\Lambda^{\alpha}(J)\right\} \label{renyi}
\end{eqnarray}
can be expressed directly.

\section{Density Matrix in the Large Block Limit}
\label{sec6}

Consider the homogeneous AKLT model with spin-$S$ at each site. In the limit $L\rightarrow\infty$ (that is when the size of the block becomes large), we learned from \cite{FK}, \cite{Had}, \cite{KHH}, \cite{VLRK} that the von Neumann entropy reaches the saturated value $S_{v.N}=\ln\left(S+1\right)^{2}$. It was also proved in \cite{XKHK} that the density matrix (denoted by $\vec{\rho}_{\infty}$ in the limit) takes the form
\begin{eqnarray}
	\vec{\rho}_{\infty}=\frac{1}{(S+1)^{2}}I_{(S+1)^{2}}\oplus \Phi_{\infty}, \label{denl}
\end{eqnarray}
where $I_{(S+1)^{2}}$ is the identity of dimension $(S+1)^{2}$ and $\Phi_{\infty}$ is an infinite dimensional matrix with only zero entries. In this section, we generalize these properties in the limiting case to the inhomogeneous model.

Let's apply the density matrix $\vec{\rho}_{L}$ (\ref{matr}) to the state $|\mbox{G}; J, \hat{\Omega}\rangle$ (\ref{eige}) and get
\begin{eqnarray}
	&&\vec{\rho}_{L}|\mbox{G}; J, \hat{\Omega}\rangle \label{appl1} \\
	&=&\frac{\displaystyle\prod^{L}_{j=0}(M_{j,j+1}+1)}{\displaystyle\prod^{L}_{j=1}(2S_{j}+1)!}\frac{1}{(4\pi)^{2}}
	\int d\hat{\Omega}_{0}d\hat{\Omega}_{L+1}B^{\dagger}|\mbox{VBS}_{L}\rangle\langle \mbox{VBS}_{L}|BA^{\dagger}_{J}|\mbox{VBS}_{L}\rangle. \nonumber
\end{eqnarray}
Using the coherent state representation (\ref{cohe}) and completeness relation (\ref{comp}), the factor $\langle \mbox{VBS}_{L}|BA^{\dagger}_{J}|\mbox{VBS}_{L}\rangle$
in (\ref{appl1}) can be re-written as
\begin{eqnarray}
	&&\langle \mbox{VBS}_{L}|BA^{\dagger}_{J}|\mbox{VBS}_{L}\rangle \label{appl2} \\
	&=&\displaystyle\left[\prod^{L}_{j=1}\frac{(2S_{j}+1)!}{4\pi}\right]\int \left(\prod^{L}_{j=1}d\hat{\Omega}_{j}\right)\prod^{L-1}_{j=1}\left[\frac{1}{2}(1-\hat{\Omega}_{j}\cdot\hat{\Omega}_{j+1})\right]^{M_{j,j+1}}
	\nonumber \\
	&&\cdot\left(u_{0}v_{1}-v_{0}u_{1}\right)^{M_{01}} \left(uu^{\ast}_{1}+vv^{\ast}_{1}\right)^{J_{-}+J}\left(u^{\ast}_{1}v^{\ast}_{L}-v^{\ast}_{1}u^{\ast}_{L}\right)^{J_{+}-J}
\nonumber \\	&&\cdot\left(uu^{\ast}_{L}+vv^{\ast}_{L}\right)^{-J_{-}+J}\left(u_{L}v_{L+1}-v_{L}u_{L+1}\right)^{M_{L,L+1}}. \nonumber
\end{eqnarray}
We plug the expression (\ref{appl2}) into (\ref{appl1}). Using transformation properties under $SU(2)$ and binomial expansion, the integral over $\hat{\Omega}_{0}$ yields that
\begin{eqnarray}
	\int d\hat{\Omega}_{0}\left(u^{\ast}_{0}b^{\dagger}_{1}-v^{\ast}_{0}a^{\dagger}_{1}\right)^{M_{01}}\left(u_{0}v_{1}-v_{0}u_{1}\right)^{M_{01}}
	=\frac{4\pi}{M_{01}+1}\left(u_{1}a^{\dagger}_{1}+v_{1}b^{\dagger}_{1}\right)^{M_{01}}. \label{int0} 
\end{eqnarray}
Similarly we can perform the integral over $\hat{\Omega}_{L+1}$. Then using expansion (\ref{expansion}) and orthogonality of spherical harmonics, other integrals over $\hat{\Omega}_{j}$ with $j=2, \ldots, L-1$ in (\ref{appl2}) can be performed. As a result, the following expression is obtained from (\ref{appl1}):
\begin{eqnarray}
\vec{\rho}_{L}|\mbox{G}; J, \hat{\Omega}\rangle =  \frac{1}{(4\pi)^{2}}\sum^{M_{<}}_{l=0}(2l+1)\left[\prod^{L-1}_{j=1}\lambda(l,M_{j,j+1})\right]
K^{\dagger}_{l}(\hat{\Omega})
\left|\mbox{VBS}_{L}\right\rangle. \label{sum}
\end{eqnarray}
The operator $K^{\dagger}_{l}(\hat{\Omega})$ involved in (\ref{sum}) is defined as
\begin{eqnarray}
	K^{\dagger}_{l}(\hat{\Omega})&\equiv&\int d\hat{\Omega}_{1} d\hat{\Omega}_{L} 
	P_{l}(\hat{\Omega}_{1} \cdot \hat{\Omega}_{L})
	\left(u_{1}a^{\dagger}_{1}+v_{1}b^{\dagger}_{1}\right)^{M_{01}} \left(uu^{\ast}_{1}+vv^{\ast}_{1}\right)^{J_{-}+J}
\label{inte} \\ 
&&\cdot \left(u^{\ast}_{1}v^{\ast}_{L}-v^{\ast}_{1}u^{\ast}_{L}\right)^{J_{+}-J}
\left(uu^{\ast}_{L}+vv^{\ast}_{L}\right)^{-J_{-}+J}
\left(u_{L}a^{\dagger}_{L}+v_{L}b^{\dagger}_{L}\right)^{M_{L,L+1}}. \nonumber
\end{eqnarray} 
It is expressed as an integral depending on the order $l$ of the Legendre polynomial $P_{l}(\hat{\Omega}_{1} \cdot \hat{\Omega}_{L})$.

There was no ambiguity in defining the large block limit in the homogeneous AKLT model \cite{XKHK}. However, in the inhomogeneous model we must specify the behavior of ending spins in the large block limit. So we define the large block limit as when $L\rightarrow\infty$, the two ending spins approach definite values, namely, $M_{01}\rightarrow S_{-}$ and $M_{L,L+1}\rightarrow S_{+}$. Then we realize from (\ref{lambdalm}) that as $L\rightarrow\infty$, $\prod^{L-1}_{j=1}\lambda(l,M_{j,j+1})\rightarrow \delta_{l,0}$. Therefore only the first term with $l=0$ is left in (\ref{sum}). So that we need only to calculate the limiting $K^{\dagger}_{0}(\hat{\Omega})$:
\begin{eqnarray}
	K^{\dagger}_{0}(\hat{\Omega})&\stackrel{L\rightarrow\infty}{\longrightarrow}&\int d\hat{\Omega}_{1} d\hat{\Omega}_{L} 
	\left(u_{1}a^{\dagger}_{1}+v_{1}b^{\dagger}_{1}\right)^{S_{-}} \left(uu^{\ast}_{1}+vv^{\ast}_{1}\right)^{J_{-}+J}    \nonumber \\ &&\cdot
\left(u^{\ast}_{1}v^{\ast}_{L}-v^{\ast}_{1}u^{\ast}_{L}\right)^{J_{+}-J}
\left(uu^{\ast}_{L}+vv^{\ast}_{L}\right)^{-J_{-}+J}
\left(u_{L}a^{\dagger}_{L}+v_{L}b^{\dagger}_{L}\right)^{S_{+}}. \label{k0}
\end{eqnarray} 
Here both $J_{-}$ and $J_{+}$ take the limiting values $\frac{1}{2}(S_{-}-S_{+})$ and $\frac{1}{2}(S_{-}+S_{+})$, respectively.

It is useful to know transformation properties of the integrand in (\ref{k0}) under $SU(2)$.
The pair of variables $(u, v)$ defined in (\ref{spin}) and bosonic annihilation operators $(a, b)$ in the Schwinger representation both transform as spinors under $SU(2)$. That is to say, if we take an arbitrary element $\vec{D}\in SU(2)$ ($2\times 2$ matrix), then $(u, v)$, \textit{etc.} transform according to
\begin{eqnarray}
\left( \begin{array}{c}
	u \\ v 
\end{array}\right)
\rightarrow \vec{D}
\left( \begin{array}{c}
	u \\ v
\end{array} \right). \label{tran}
\end{eqnarray}
On the other hand, $(u^{\ast}, v^{\ast})$, $(-v, u)$, $(a^{\dagger}, b^{\dagger})$ and $(-b, a)$ transform conjugately to $(u, v)$. That is to say $(u^{\ast}, v^{\ast})$, \textit{etc.} transform according to
\begin{eqnarray}
	\left( \begin{array}{c}
	u^{\ast} \\ v^{\ast} 
\end{array}\right)
\rightarrow \vec{D}^{\ast}
\left( \begin{array}{c}
	u^{\ast} \\ v^{\ast}
\end{array}\right). \label{ctra}
\end{eqnarray}
The integrand appeared in $K^{\dagger}_{0}(\hat{\Omega})$ (\ref{k0}), as well as $A^{\dagger}_{J}$ in (\ref{aope}), boundary operator $B^{\dagger}$ in (\ref{bope1}), \textit{etc.} all transform covariantly under $SU(2)$, i.e. those expressions keep their form in the new (transformed) coordinates. 

These transformation properties (\ref{tran}), (\ref{ctra}) can be used to simplify the $K^{\dagger}_{0}(\hat{\Omega})$ integral (see \cite{XKHK} for more details). We first make a $SU(2)$ transform
\begin{eqnarray}
\vec{D}_{u_{L}}=
\left( \begin{array}{cc}
	u^{\ast}_{L} & v^{\ast}_{L} \\
	-v_{L} & u_{L}
\end{array} \right), \qquad
\vec{D}_{u_{L}}
\left( \begin{array}{c}
	u_{L} \\ v_{L}
\end{array} \right)=\left( \begin{array}{c}
	1 \\ 0
\end{array} \right), \label{utra}
\end{eqnarray}
under the part of the integral (\ref{k0}) over $\hat{\Omega}_{1}$. Then after integrating over $\hat{\Omega}_{1}$ by a binomial expansion, we make an inverse transform in using $\vec{D}^{-1}_{u_{L}}=\vec{D}^{\dagger}_{u_{L}}$, consequently (\ref{k0}) is put in a form with a single integral over $\hat{\Omega}_{L}$ remaining:
\begin{eqnarray}
K^{\dagger}_{0}(\hat{\Omega})&=&\frac{4\pi}{S_{-}+1}\left(ua^{\dagger}_{1}+vb^{\dagger}_{1}\right)^{J_{-}+J} \label{rema} \\
&\cdot&\int d\hat{\Omega}_{L} 
\left(a^{\dagger}_{1}v^{\ast}_{L}-b^{\dagger}_{1}u^{\ast}_{L}\right)^{J_{+}-J}
\left(uu^{\ast}_{L}+vv^{\ast}_{L}\right)^{-J_{-}+J}   \left(u_{L}a^{\dagger}_{L}+v_{L}b^{\dagger}_{L}\right)^{S_{+}}. \nonumber
\end{eqnarray}
Similarly, we make another $SU(2)$ transform using
\begin{eqnarray}
\vec{D}_{u}=
\left( \begin{array}{cc}
	u^{\ast} & v^{\ast} \\
	-v & u
\end{array} \right), \qquad
\vec{D}_{u}
\left( \begin{array}{c}
	u \\ v
\end{array} \right)=\left( \begin{array}{c}
	1 \\ 0
\end{array} \right)
\end{eqnarray}
and integrate over $\hat{\Omega}_{L}$. At last we make an inverse transform using $\vec{D}^{-1}_{u}=\vec{D}^{\dagger}_{u}$, the final form is
\begin{eqnarray}
K^{\dagger}_{0}(\hat{\Omega})
=\frac{(4\pi)^{2}}{(S_{-}+1)(S_{+}+1)}A^{\dagger}_{J}. \label{fina}
\end{eqnarray}
This expression states that $\{|\mbox{G}; J, \hat{\Omega}\rangle\}$ is a set of eigenvectors of the density matrix as $L\rightarrow\infty$. Let's denote the density matrix in the limit by $\vec{\rho}_{\infty}$. Then (\ref{fina}) leads to the result (see (\ref{sum}))
\begin{eqnarray}
	\vec{\rho}_{\infty}|\mbox{G}; J, \hat{\Omega}\rangle=
	\frac{1}{(S_{-}+1)(S_{+}+1)}|\mbox{G}; J, \hat{\Omega}\rangle. \label{appll}
\end{eqnarray}

We find from (\ref{appll}) that the limiting eigenvalue $\Lambda_{\infty}=\frac{1}{(S_{-}+1)(S_{+}+1)}$ is independent of $J$. Any vector of the $(S_{-}+1)(S_{+}+1)$-dimensional subspace spanned by the set $\{|\mbox{G}; J, \hat{\Omega}\rangle\}$ is an eigenvector of $\vec{\rho}_{\infty}$ with the same eigenvalue $\frac{1}{(S_{-}+1)(S_{+}+1)}$. Therefore $\vec{\rho}_{\infty}$ acts on this subspace as (proportional to) the identity $I_{(S_{-}+1)(S_{+}+1)}$. So that we have proved explicitly that the density matrix takes the form 
\begin{eqnarray}
	\vec{\rho}_{\infty}=\frac{1}{(S_{-}+1)(S_{+}+1)}I_{(S_{-}+1)(S_{+}+1)}\oplus \Phi_{\infty}, \label{limitrho}
\end{eqnarray}
which is a generalization of (\ref{denl}) in the large block limit. In addition, we also derive from the eigenvalues that the von Neumann entropy $S_{v.N}=-\sum^{J_{+}}_{J=|J_{-}|}(2J+1)\Lambda_{\infty}\ln\Lambda_{\infty}$ coincides with the Renyi entropy $S_{R}=\frac{1}{1-\alpha}\ln\left\{\sum^{J_{+}}_{J=|J_{-}|}(2J+1)\Lambda^{\alpha}_{\infty}\right\}$ and is equal to the saturated value $\ln\left[(S_{-}+1)(S_{+}+1)\right]$.

\section{Conclusion}
\label{sec7}

We have studied the density matrix $\vec{\rho}_{L}$ of a block of $L$ contiguous bulk spins as a subsystem in the inhomogeneous AKLT model. The Hamiltonian is given by (\ref{hami}), which has a unique ground state under condition (\ref{condition}). The ground state is described by the VBS state (\ref{vbs}) in the Schwinger representation. The density matrix $\vec{\rho}_{L}$ (\ref{matr}) of the block is obtained by taking trace (\ref{trac}) of all spin degrees of freedom outside the block. The structure of the density matrix has been investigated both for finite and infinite blocks. It has been shown that the structure and properties of the density matrix are all generalizable to the inhomogeneous model, as conjectured in \cite{XKHK}.

For finite block, two mathematically rigorous results have been established as \textbf{Theorem 1} and \textbf{2}. In \textbf{Theorem 1} we constructed eigenvectors of the density matrix with non-zero eigenvalues. These eigenvectors $|\mbox{G}; J, \hat{\Omega}\rangle$ defined in (\ref{eige}), or $|\mbox{VBS}_{L}(J,M)\rangle$ defined in (\ref{devb}) equivalently, are proved to be the $(M_{01}+1)(M_{L,L+1}+1)$ zero-energy ground states of the block Hamiltonian (\ref{blockhamire}). Using orthogonal basis $\{|\mbox{VBS}_{L}(J,M)\rangle\}$, in \textbf{Theorem 2} an explicit expression (\ref{eivaex}) for corresponding eigenvalues in terms of Wigner $3j$-symbols is derived. Non-zero eigenvalues $\Lambda(J)$ with $J=|J_{-}|, |J_{-}|+1, \ldots, J_{+}$ depend only on $J$ and is independent of $\hat{\Omega}$ and/or $M$ in defining eigenvectors. The density matrix (\ref{fipr}) is a projector onto the subspace of dimension $(M_{01}+1)(M_{L,L+1}+1)$ spanned by the set of eigenvectors $\{|\mbox{G}; J, \hat{\Omega}\rangle\}$ and/or $\{|\mbox{VBS}_{L}(J,M)\rangle\}$.

In the large block limit $L\rightarrow\infty$, ending spins approach definite values, namely, $M_{01}\rightarrow S_{-}$ and $M_{L,L+1}\rightarrow S_{+}$. All non-zero eigenvalues $\Lambda_{\infty}$ become the same (\ref{appll}). The infinite dimensional density matrix $\vec{\rho}_{\infty}$ (\ref{limitrho}) is a projector onto a $(S_{-}+1)(S_{+}+1)$-dimensional subspace in which it is proportional to the identity. The von Neumann entropy $S_{v.N}$ coincides with the Renyi entropy $S_{R}$ and is equal to the saturated value $\ln\left[(S_{-}+1)(S_{+}+1)\right]$. In the limit the Renyi entropy is $\alpha$ independent, which behaves quite differently from the XY model where the Renyi entropy has an essential singularity as a function of $\alpha$ (see \cite{IJK}, \cite{FIJK}, \cite{FIK}).

\begin{acknowledgements}
The authors would like to thank Professor Heng Fan, Professor Anatol N. Kirillov and Professor Sergey Bravyi for valuable discussions and suggestions. The work is supported by NSF Grant DMS-0503712 and the Japan Society for the Promotion of Science.
\end{acknowledgements}

\appendix

\section{Orthogonality of Degenerate VBS States}
\label{secA}

The set of degenerate VBS states $\{|\mbox{VBS}_L(J,M)\rangle, J=|J_{-}|,...,J_{+}, M=-J,...,J\}$ introduced in (\ref{devb}) are mutually orthogonal. To show this, 
it is convenient to introduce the total spin operators of the subsystem:
\begin{equation}
S^{+}_{\mbox{\scriptsize{tot}}}=\sum^{L}_{j=1} a^{\dagger}_{j} b_{j}, \qquad
S^{-}_{\mbox{\scriptsize{tot}}}=\sum^{L}_{j=1} b^{\dagger}_{j} a_{j}, \qquad
S^{z}_{\mbox{\scriptsize{tot}}}=\sum^{L}_{j=1} \frac{1}{2}(a^{\dagger}_{j} a_{j}-b^{\dagger}_{j} b_{j}). \label{totalspinoperator}
\end{equation}
First we show that the set of operators $\{ S^{+}_{\mbox{\scriptsize{tot}}}, S^{-}_{\mbox{\scriptsize{tot}}}, S^{z}_{\mbox{\scriptsize{tot}}} \}$ commute with the product of valence bonds, i.e.
\begin{equation}
[S^{\pm}_{\mbox{\scriptsize{tot}}}, \prod^{L-1}_{j=1} (a^{\dagger}_{j} b^{\dagger}_{j+1}-b^{\dagger}_{j} a^{\dagger}_{j+1})^{M_{j,j+1}}]=0, \quad 
[S^{z}_{\mbox{\scriptsize{tot}}}, \prod^{L-1}_{j=1} (a^{\dagger}_{j} b^{\dagger}_{j+1}-b^{\dagger}_{j} a^{\dagger}_{j+1})^{M_{j,j+1}} ]=0.
\label{commute}
\end{equation}
These commutation relations (\ref{commute}) can be shown in similar ways. Take the commutator with $S^{+}_{\mbox{\scriptsize{tot}}}$ first. We re-write the commutator as
\begin{eqnarray}
&&[S^{+}_{\mbox{\scriptsize{tot}}}, \prod_{j=1}^{L-1} (a^{\dagger}_{j} b^{\dagger}_{j+1}-b^{\dagger}_{j} a^{\dagger}_{j+1})^{M_{j,j+1}}] \nonumber \\
&=& \sum_{j=1}^{L-1} (a^{\dagger}_{1} b^{\dagger}_{2}-b^{\dagger}_{1} a^{\dagger}_{2})^{M_{12}} \cdots 
[S^{+}_{\mbox{\scriptsize{tot}}}, (a^{\dagger}_{j} b^{\dagger}_{j+1}-b^{\dagger}_{j} a^{\dagger}_{j+1})^{M_{j,j+1}}] \cdots 
\nonumber \\
&&\cdots(a^{\dagger}_{L-1} b^{\dagger}_{L}-b^{\dagger}_{L-1} a^{\dagger}_{L})^{M_{L-1,L}} \nonumber \\
&=& \sum_{j=1}^{L-1} (a^{\dagger}_{1} b^{\dagger}_{2}-b^{\dagger}_{1} a^{\dagger}_{2})^{M_{12}} \cdots 
[S^{+}_{j} + S^{+}_{j+1}, (a^{\dagger}_{j} b^{\dagger}_{j+1}-b^{\dagger}_{j} a^{\dagger}_{j+1})^{M_{j,j+1}}] \cdots 
\nonumber \\
&&\cdots(a^{\dagger}_{L-1} b^{\dagger}_{L}-b^{\dagger}_{L-1} a^{\dagger}_{L})^{M_{L,L+1}}. \nonumber \\
\end{eqnarray}
Then using commutators $[a_{i}, a^{\dagger}_{j}]=\delta_{ij}$ and $[b_{i}, b^{\dagger}_{j}]=\delta_{ij}$, we find that
\begin{eqnarray}
&& [S^{+}_{j} + S^{+}_{j+1}, (a^{\dagger}_{j} b^{\dagger}_{j+1}-b^{\dagger}_{j} a^{\dagger}_{j+1})^{M_{j,j+1}}]
\nonumber \\
&=& [a^{\dagger}_{j} b_{j}+a^{\dagger}_{j+1} b_{j+1}, (a^{\dagger}_{j} b^{\dagger}_{j+1}-b^{\dagger}_{j} a^{\dagger}_{j+1})^{M_{j,j+1}}] \nonumber \\
&=& a^{\dagger}_{j} [b_{j}, (a^{\dagger}_{j} b^{\dagger}_{j+1}-b^{\dagger}_{j} a^{\dagger}_{j+1})^{M_{j,j+1}}]+
      a^{\dagger}_{j+1} [b_{j+1}, (a^{\dagger}_{j} b^{\dagger}_{j+1}-b^{\dagger}_{j} a^{\dagger}_{j+1})^{M_{j,j+1}}]
      \nonumber \\
&=& a^{\dagger}_{j} (-M_{j,j+1})a^{\dagger}_{j+1} (a^{\dagger}_{j} b^{\dagger}_{j+1}-b^{\dagger}_{j} a^{\dagger}_{j+1})^{M_{j,j+1}-1}
\nonumber \\
    &&+a^{\dagger}_{j+1} M_{j,j+1} a^{\dagger}_{j} (a^{\dagger}_{j} b^{\dagger}_{j+1}-b^{\dagger}_{j} a^{\dagger}_{j+1})^{M_{j,j+1}-1}
\nonumber \\
&=&0. \label{comm}
\end{eqnarray}
Therefore $[S^{+}_{\mbox{\scriptsize{tot}}}, \prod_{j=1}^{L-1} (a^{\dagger}_{j} b^{\dagger}_{j+1}-b^{\dagger}_{j} a^{\dagger}_{j+1})^{M_{j,j+1}}]=0$. In (\ref{comm}) we have used \\ $[b_{j}, (a^{\dagger}_{j} b^{\dagger}_{j+1}-b^{\dagger}_{j} a^{\dagger}_{j+1})^{M_{j,j+1}}]=-M_{j,j+1} a^{\dagger}_{j+1} (a^{\dagger}_{j} b^{\dagger}_{j+1}-b^{\dagger}_{j} a^{\dagger}_{j+1})^{M_{j,j+1}-1}$. 
In a parallel way, we find that the commutator with $S^{-}_{\mbox{\scriptsize{tot}}}$ also vanishes. Next we consider the commutator with $S^{z}_{\mbox{\scriptsize{tot}}}$:
\begin{eqnarray}
&&[S^{z}_{\mbox{\scriptsize{tot}}}, \prod_{j=1}^{L-1} (a^{\dagger}_{j} b^{\dagger}_{j+1}-b^{\dagger}_{j} a^{\dagger}_{j+1})^{M_{j,j+1}}] \nonumber \\
&=& \sum_{j=1}^{L-1} (a^{\dagger}_{1} b^{\dagger}_{2}-b^{\dagger}_{1} a^{\dagger}_{2})^{M_{12}} \cdots 
[S^{z}_{j} + S^{z}_{j+1}, (a^{\dagger}_{j} b^{\dagger}_{j+1}-b^{\dagger}_{j} a^{\dagger}_{j+1})^{M_{j,j+1}}] \cdots 
\nonumber \\
&&\cdots(a^{\dagger}_{L-1} b^{\dagger}_{L}-b^{\dagger}_{L-1} a^{\dagger}_{L})^{M_{L-1,L}}. \nonumber \\
\label{comSz}
\end{eqnarray}
In the right hand side of (\ref{comSz}), the commutator involved also vanishes because
\begin{eqnarray}
&&[S^{z}_{j} + S^{z}_{j+1}, (a^{\dagger}_{j} b^{\dagger}_{j+1}-b^{\dagger}_{j} a^{\dagger}_{j+1})^{M_{j,j+1}}] 
\nonumber \\
&=& \frac{1}{2}[a^{\dagger}_{j} a_{j} -b^{\dagger}_{j} b_{j} +a^{\dagger}_{j+1} a_{j+1} -b^{\dagger}_{j+1} b_{j+1}, (a^{\dagger}_{j} b^{\dagger}_{j+1}-b^{\dagger}_{j} a^{\dagger}_{j+1})^{M_{j,j+1}}] 
\nonumber \\
&=& a^{\dagger}_{j}  [a_{j}, (a^{\dagger}_{j} b^{\dagger}_{j+1}-b^{\dagger}_{j} a^{\dagger}_{j+1})^{M_{j,j+1}}]
     -b^{\dagger}_{j}  [b_{j} ,(a^{\dagger}_{j} b^{\dagger}_{j+1}-b^{\dagger}_{j} a^{\dagger}_{j+1})^{M_{j,j+1}}] 
\nonumber \\
&+& a^{\dagger}_{j+1} [a_{j+1}, (a^{\dagger}_{j} b^{\dagger}_{j+1}-b^{\dagger}_{j} a^{\dagger}_{j+1})^{M_{j,j+1}}]
     -b^{\dagger}_{j+1} [b_{j+1}, (a^{\dagger}_{j} b^{\dagger}_{j+1}-b^{\dagger}_{j} a^{\dagger}_{j+1})^{M_{j,j+1}}]
\nonumber \\
&=& 0  \label{com1}
\end{eqnarray}
Substituting (\ref{com1}) into (\ref{comSz}), we obtain $[S^{z}_{\mbox{\scriptsize{tot}}}, \prod_{j=1}^{L-1} (a^{\dagger}_{j} b^{\dagger}_{j+1}-b^{\dagger}_{j} a^{\dagger}_{j+1})^{M_{j,j+1}}]=0$.
Now we shall show that the state $|\mbox{VBS}_L(J,M)\rangle$ is an eigenstate of $S^{z}_{\mbox{\scriptsize{tot}}}$ and the square of the total spin $\vec{S}^2_{\mbox{\scriptsize{tot}}}=\frac{1}{2}(S^{+}_{\mbox{\scriptsize{tot}}}S^{-}_{\mbox{\scriptsize{tot}}}+S^{-}_{\mbox{\scriptsize{tot}}}S^{+}_{\mbox{\scriptsize{tot}}}) + (S^{z}_{\mbox{\scriptsize{tot}}})^{2}$ with eigenvalues $M$ and $J(J+1)$, respectively. 
Using the commutation relations (\ref{commute}), we can show that
\begin{eqnarray}
S^{\pm}_{\mbox{\scriptsize{tot}}}|\mbox{VBS}_L(J,M)\rangle = \prod_{j=1}^{L-1} (a^{\dagger}_{j} b^{\dagger}_{j+1}-b^{\dagger}_{j} a^{\dagger}_{j+1})^{M_{j,j+1}} (S^{\pm}_{1} + S^{\pm}_{L})|J,M \rangle_{1,L} |\mbox{vac}\rangle_{2, ..., L-1} \nonumber \\
S^{z}_{\mbox{\scriptsize{tot}}}|\mbox{VBS}_L(J,M)\rangle = \prod_{j=1}^{L-1} (a^{\dagger}_{j} b^{\dagger}_{j+1}-b^{\dagger}_{j} a^{\dagger}_{j+1})^{M_{j,j+1}} (S^{z}_{1} + S^{z}_{L})|J,M \rangle_{1,L} |\mbox{vac}\rangle_{2,..., L-1}. \nonumber \\
\end{eqnarray}
Then from the definition of the state $|\mbox{VBS}_L(J,M)\rangle$ and the following relations:
\begin{eqnarray}
(S^{+}_1+S^{+}_{L})|J,M \rangle_{1,L}&=& \sqrt{(J \mp M) (J \pm M +1)} |J, M\pm1 \rangle, 
\nonumber \\
(S^{z}_{1}+S^{z}_{L})|J,M \rangle_{1,L}\rangle &=& M |J,M \rangle_{1,L},
\end{eqnarray}
we obtain 
\begin{eqnarray}
S^{\pm}_{\mbox{\scriptsize{tot}}}|\mbox{VBS}_L(J,M)\rangle &=& \sqrt{(J \mp M) (J \pm M +1)} |\mbox{VBS}_L(J,M\pm1)\rangle, \nonumber \\
S^z_{\mbox{\scriptsize{tot}}}|\mbox{VBS}_L(J,M)\rangle &=& M |\mbox{VBS}_L(J,M)\rangle
\end{eqnarray}
and hence $\vec{S}^{2}_{\mbox{\scriptsize{tot}}}|\mbox{VBS}_L(J,M)\rangle=J(J+1)|\mbox{VBS}_L(J,M)\rangle$.
It is now proved that $|\mbox{VBS}_L(J,M)\rangle$ is an eigenstate of $S^{z}_{\mbox{\scriptsize{tot}}}$ and $\vec{S}^{2}_{\mbox{\scriptsize{tot}}}$ with eigenvalues $M$ and $J(J+1)$, respectively. Therefore the states with different eigenvalues $(J, M)$ are orthogonal to each other.

With the introduction of total spin operators of the block (\ref{totalspinoperator}), it is straightforward to prove that normalization of the degenerate VBS state $|{\rm VBS}_L(J,M)\rangle$ depends only on $J$ and is independent of $M$. We prove the statement as follows:
\begin{eqnarray}
&& \langle \mbox{VBS}_{L}(J,M\pm 1)|\mbox{VBS}_{L}(J,M\pm 1)\rangle
\nonumber \\
&=& \frac{1}{(J\mp M)(J \pm M+1)}\langle\mbox{VBS}_{L}(J,M)|S^{\mp}_{\mbox{\scriptsize{tot}}}S^{\pm}_{\mbox{\scriptsize{tot}}}|\mbox{VBS}_{L}(J,M)\rangle
\nonumber \\
&=& \frac{1}{(J\mp M)(J \pm M+1)}\langle\mbox{VBS}_{L}(J,M)|(\vec{S}^{2}_{\mbox{\scriptsize{tot}}}-(S^{z}_{\mbox{\scriptsize{tot}}})^{2}\mp S^{z}_{\mbox{\scriptsize{tot}}})|\mbox{VBS}_{L}(J,M)\rangle
\nonumber \\
&=& \langle\mbox{VBS}_{L}(J,M)|\mbox{VBS}_{L}(J,M)\rangle.
\end{eqnarray}
Here we have used the fact that $|\mbox{VBS}_{L}(J,M)\rangle$ is the eigenstate of $\vec{S}^{2}_{\mbox{\scriptsize{tot}}}$ and $S^{z}_{\mbox{\scriptsize{tot}}}$ with eigenvalues $J(J+1)$ and $M$, respectively.

\end{document}